\documentclass[%
 aps,
 prb,%
 amsmath,amssymb,
reprint,%
superscriptaddress,showpacs]{revtex4-1}

\usepackage{color, graphicx}% Include figure files
\usepackage{dcolumn}% Align table columns on decimal point
\usepackage{bm}% bold math
\usepackage{amssymb}
\usepackage{latexsym}
\usepackage{amsfonts}
\usepackage{amsmath}
\usepackage{multirow}
\usepackage{ifthen}

\begin{document}

%\preprint{}

%Title of paper
\title{Enhancement of sound in chirped sonic crystals}

\author{V. Romero-Garc\'ia}
\author{R. Pic\'o}
\author{A. Cebrecos}
\author{V. J. S\'anchez-Morcillo}
 \affiliation{Instituto de Investigaci\'on para la Gesti\'on Integrada de zonas Costeras, Universitat Polit\`ecnica de Val\`encia, Paranimf 1, 46730, Grao de Gandia, Val\`encia, Spain}
 
\author{K. Staliunas} 
\affiliation{ICREA, Departament de F\'isica i Enginyeria Nuclear, Universitat Polit\`ecnica de Catalunya, Colom 11, E-08222 Terrasa, Barcelona, Spain}

%\date{\today}

\begin{abstract}
We propose and experimentally demonstrate a novel mechanism of sound wave concentration based on soft reflections in chirped sonic crystals. The reported controlled field enhancement occurs at around particular (bright) planes in the crystal, and is related to a progressive slowing down of the sound wave as it propagates along the material. At these bright planes, a substantial concentration of the energy (with a local increase up to 20 times) was obtained for a linear chirp and for frequencies around the first band gap. A simple couple mode theory is proposed, that interprets and estimates the observed effects. The results are obtained for the case of sound waves and sonic crystals, however they are extendable to other type of waves in modulated host matter.
\end{abstract}

% insert suggested PACS numbers in braces on next line
\pacs{43.20.Fn, 43.20.Gp, 43.20.Mv, 63.20.-e}
% insert suggested keywords - APS authors don't need to do this
%\keywords{Phononic crystal, Sonic Crystal, Complex band structures, Evanescent modes, Waveguides}

%\maketitle must follow title, authors, abstract, \pacs, and \keywords
\maketitle

% body of paper here - Use proper section commands
% References should be done using the \cite, \ref, and \label commands
Manipulation and control of wave propagation, a problem of fundamental interest, is at root of many applications in different branches of science and technology. One important issue of wave manipulation is the localization and concentration (or local enhancement) of the wave energy. Artificial materials, and among them, artificial crystals are emerging as promising tools for manipulating wave propagation. In the case of sound waves considered here, such artificial periodic materials are called sonic crystals, structurally similar to photonic crystals in the field of optics. They are synthetic materials formed by a periodic distribution of elements or scatterers, whose properties (i.e., elasticity and density) differ from those of the host medium. This results in a periodic modulation of the acoustic properties of the medium at the scale of wavelength. The strong interest in these materials comes from their ability of manipulating the propagation of sound waves, due to their peculiar dispersive properties. A number of exotic and useful effects such as the formation of band-gaps, \cite{Kushwaha93, Martinez95} negative refraction,\cite{Zhang04} birefraction,\cite{Lu07} self-collimation,\cite{Espinosa07} extraordinary transmission,\cite{Zhou10} or local resonances \cite{Liu00a} have been so far demonstrated for sound waves. Utilizing these wave propagation effects, novel devices such as acoustic frequency filters,\cite{Khelif03} spatial (angular) filters,\cite{Pico12} lenses,\cite{Cervera02} or diodes\cite{Li11} have been proposed and demonstrated.

We present here a novel wave propagation effect, consisting in specifically the wave energy concentration due to progressive decrease of the group velocity in chirped sonic crystals, in which the lattice constant, i.e. the distance between scatterers in longitudinal (the wave propagation) direction, gradually changes along the propagation direction. We propose and demonstrate here a substantial increase of the wave intensity in controlled zones inside the crystal. Chirped (sometimes called graded) crystals have been introduced in optics \cite{Cassan11} and acoustics \cite{Kushwaha98, Psarobas02, Wu11} for different purposes, such as opening wide full band gaps or waveguiding of beams. An intriguing phenomenon shown in chirped crystals is the smooth deflection of a light beam from the straight trajectory as it propagates through the crystal, the so-called mirage effect \cite{Centeno06}. 

Another interesting effect reported recently is the so-called rainbow trapping effect, the dependence of the turning point position on the color of radiation. It has been predicted for one-dimensionally modulated chirped photonic structures \cite{Shen11} and tapered optical and plasmonic waveguides \cite{Smolyaninova10, Jang11}. Rainbow trapping and wave enhancement are two different physical effects (the latter occurs even for monochromatic radiation), although they may occur simultaneously in chirped structures when the incident radiation is broadband. In this letter, in addition to the extraordinary sound wave enhancement effect, which is the main result reported, we also present a “sound rainbow” trapping effect for acoustic waves as a secondary result.

\begin{figure}
\begin{center}
\includegraphics[width=85mm]{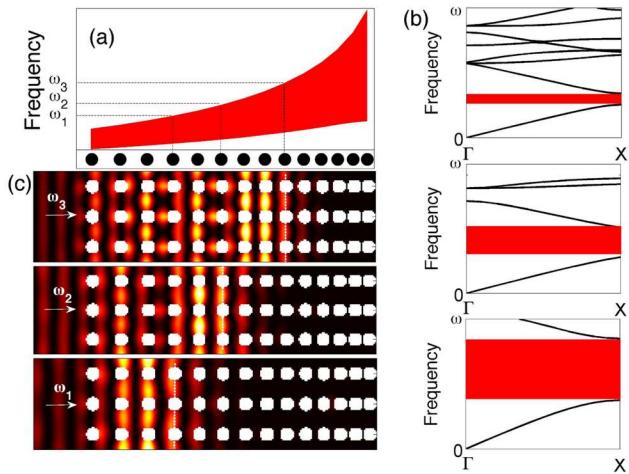}
\end{center} %\begin{quote}
\caption{(Color online) (a) Dependence of the local band gaps on the local lattice constant along the chirped sonic crystal. (b) Band structure (local dispersion curves) evaluated at different depths: at the entrance (top), at the middle (center) and at the exit (bottom) of the sonic crystal. (c) Intensity of the acoustic field calculated using Multiple Scattering technique inside the chirped structure for the frequencies in (a) and (b).}
 \label{fig:fig1}
%\end{quote} 
\end{figure}

Wave reflection from a band-gap in a chirped structure is peculiar. The dispersion curves $\omega(k)$ at- and close to the band-edges, develop nearly horizontal segments, which corresponds to small or zero group velocity of the wave, since $v_g=\partial \omega/\partial k$. The occurrence of the controlled sound enhancement requires that the crystal at the entrance plane be within the transparency range for the incoming wave, whose frequency is above the first band-gap, as shown in Fig. \ref{fig:fig1}. Note that here, not the wave frequency but the central (Bragg) frequency and the width of the band-gap is considered variable along the structure. Figures \ref{fig:fig1}(a) and \ref{fig:fig1}(b) show the variation of the band-gaps along the chirped sonic crystal, and the local dispersion curves at different depths, respectively. By local dispersion relation we mean the dispersion of an infinitely extended periodic crystal, for parameters (lattice constant, filling factor) corresponding to a particular depth of the chirped crystal. The wave entering into the crystal is gradually slowing down, as the “local” band-gaps are approaching the wave frequency in the course of propagation. Finally at a particular depth corresponding to the band-edge, the wave, literally speaking, stops, turns around, and starts propagating back. In other words it experiences a “soft” reflection. This effect is demonstrated in Fig. \ref{fig:fig1}(c), which shows the wave propagating through the crystal as obtained by numerical simulation using the multiple scattering theory approach \cite{Martin06, Chen01}. The frequencies of the incident waves in the simulations correspond to local band-gaps at different depths. Figure \ref{fig:fig1}(c) evidences that the intensity of the wave increases substantially in the soft reflection area. Most importantly and in opposition to the case of perfectly periodic crystals (constant lattice period) in which only some discrete frequencies can be enhanced by the Fabry-P\'erot resonances, chirped crystals can localize the energy for a wide range of frequencies in a controlled way by the gradually change of the lattice constant. 

\begin{figure}
\begin{center}
\includegraphics[width=85 mm]{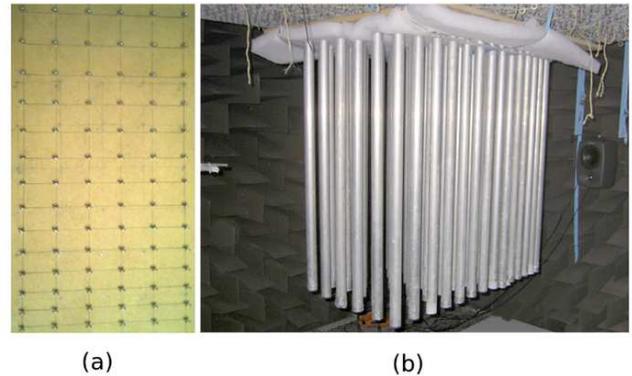}
\end{center} %\begin{quote}
\caption{(Color online) Photographs of the experimental setup. (a) the grid of hanging points of the cylinders. (b) the chirped sonic crystal hunging vertically in the anechoic chamber.}
 \label{fig:fig2}
%\end{quote} 
\end{figure}

An experimental setup was designed to demonstrate the predicted extraordinary enhancement effect, and to obtain quantitative data of the acoustic field inside the structure. It consists in a two-dimensional sonic crystal with rectangular local symmetry, as illustrated in Fig. \ref{fig:fig2}, made of acoustically rigid aluminum cylinders, of radius $r = 2$ cm, embedded in air. The spatial period is constant in transverse-to-propagation direction $y$, $a_y =10$ cm, while a longitudinal chirp is introduced in the period along the propagation direction $x$. The adimensional chirp parameter is defined as $\alpha=(a_j-a_{j+1})/a_j$, where $a_j$ is the local longitudinal lattice constant at $j$-th layer. For our particular crystal case $a_0=10$ cm (initial period), $a_{13}=4.8$ cm (final period), and a gradient $\alpha=0.055$. The sign of the chirp can be either positive or negative, corresponding to lattice constant decreasing or increasing along the propagation direction. In case of identical scatterers, as used in our study, the filling fraction for the positive (negative) chirped structures increases (decreases) in the propagation direction. This has a consequence of broadening of the local bandgap shown in Fig. \ref{fig:fig1}(a).

The measurements were performed in an echo-free chamber with dimensions $8\times 6\times 3$ m$^3$, using the automatized acquisition system 3DReAMS (3D Robotized e-Acoustic Measurement System).\cite{Romero10b} This system enables the measurement of pressure fields along complicated trajectories as well as inside the crystals. Fig. \ref{fig:fig2}(a) shows the grid of hanging points of the cylinders used to design the chirped structure. Fig. \ref{fig:fig2}(b) is a photograph showing the source (a loudspeaker) and the crystal inside the chamber (for propagation directed downwards in Fig. \ref{fig:fig2}(a)).

\begin{figure}
\begin{center}
\includegraphics[width=85mm]{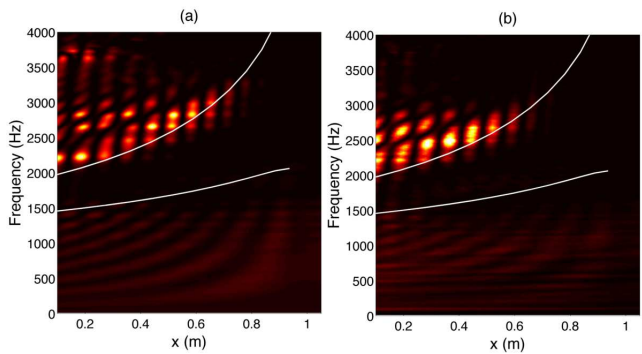}
\end{center} %\begin{quote}
\caption{(Color online) Sound distribution inside the crystal, along the x-axis, for varying frequencies (a) Multiple Scattering simulation and (b) experiment. White continuous lines show the spatially varying edges of local band-gaps. Color scale represents the values of the acoustic intensity.}
 \label{fig:fig3}
%\end{quote} 
\end{figure}

\begin{figure}
\begin{center}
\includegraphics[width=85mm]{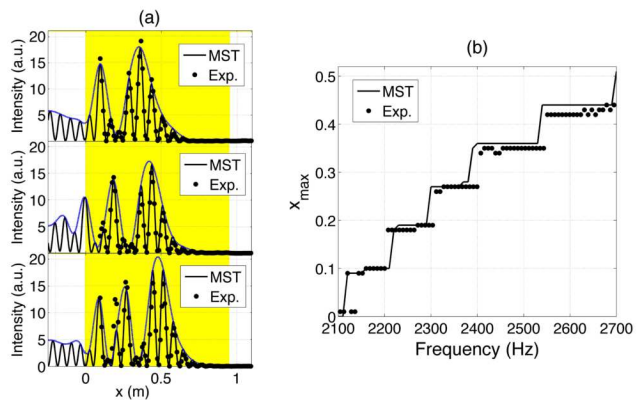}
\end{center} %\begin{quote}
\caption{(Color online) (a) Numerical simulation results (continuous line) and experimental results (dots) for the acoustic intensity at the central section inside the crystal for the three frequencies: 2500 Hz, 2600 Hz, and 2700 Hz. The shaded (yellow) rectangle denotes the area covered by the crystal. (b) Position of the maximum value of concentration of energy inside the crystal depending on the frequency of the incident wave.}
 \label{fig:fig4}
%\end{quote} 
\end{figure}

The experimental measurements are in excellent agreement with numerical calculations, as shown in Figs. \ref{fig:fig3} and \ref{fig:fig4}. We recorded the sound intensity profile along a fixed transverse position by translating the microphone along the x-axis through the void space between the rows of scatterers. In this way, we obtained two-dimensional space-frequency plots as shown in Fig. \ref{fig:fig3}, from numerical (a) and experimental (b) data. White continuous lines mark the positions of the boundaries of the first band gap. Note the concentration of acoustic energy at positions corresponding to just before the upper band edge. 

Figure \ref{fig:fig4}(a) shows the axial distributions obtained experimentally (dots) and theoretically (continuous lines) for three particular frequencies. In both cases small-scale fringes are observed, corresponding to the local Bloch mode, as well as a large-scale oscillations or envelope (dashed line) of the Bloch mode, to be discussed below. Figure \ref{fig:fig4}(b) represents the theoretical calculation of the position of the maximum value of concentration of acoustic energy inside the crystal depending on the frequency. In correspondence with the results in Fig. \ref{fig:fig1}, the position of the maximal energy concentration shifts deeper into the bulk of the structure, as the frequency is increased. Note also that, since the incident amplitude was normalized to unity, at the maximum value, the intensity has been recorded up to around 20 times higher than incident. For usual reflection between two different homogeneous media or from a purely band-gap material in the range of the band-gap, only an increase of 4 times of the local intensity is possible (as the interference pattern is formed from forward and fully reflected backward wave). For the case of periodic structures the wave penetrates into the reflecting material evanescently, i.e. with exponential decay \cite{Romero10b} and never shows an increase of intensity.

The increase of the intensity field observed in Fig. \ref{fig:fig4}, can be understood from a coupling between the forward and backward waves inside the chirped crystal. In order to interpret the results, we propose a simple coupled mode analytical theory for the propagation of acoustic plane waves inside a one-dimensional chirped crystal (in optics also known as chirped mirror). This dimension reduction is possible because the first band gap in $\Gamma$X direction essentially appears due to a resonant coupling between the forward and the backward waves. The contributions of the wave components propagating to transverse directions are negligible. Assuming that the full pressure field consists of forward and backward propagating waves, $P=A(x)e^{\imath k x-\imath \omega t} + B(x)e^{-\imath k x-\imath \omega t}+c.c.$ the following coupled amplitude equations can be systematically obtained from wave equations,
\begin{eqnarray}
\frac{dA}{dx}=\imath\frac{\sqrt{s}}{a(x)}Be^{2\imath\Delta q(x) x},\nonumber\\
\frac{dB}{dx}=-\imath\frac{\sqrt{s}}{a(x)}Ae^{-2\imath\Delta q(x) x},
\label{eq:eq1}
\end{eqnarray}
where $s$ is the back-reflection coefficient by one row of scatterers, $a(x)$ the variable longitudinal period and $\Delta q(x)=2\pi/\lambda-\pi/a(x)$ is the detuning from the Bragg frequency.

From the numerical study of the scattering by only one row of the structure, we estimate that the back reflected intensity is around 40\% of the incident, so $s\simeq0.4$. The same numerical study reveals that the scattering into field components propagating at transverse direction is only around 5\%, which justifies the followed one-dimensional approach, neglecting transverse modulations in the vicinity of the first bandgap. We notice that the detuning from the Bragg resonance $\Delta q(x)$ is a function of the longitudinal position x for chirped crystals. Recall that in our study the chirp is linear, given by $a(x)=a_0+\alpha(x-x_0)$.
 
Equations (\ref{eq:eq1}) can be rewritten in canonical form as
\begin{eqnarray}
\frac{d^2A}{dX^2}=\imath \epsilon(x)\frac{dA}{dX}+A,
\label{eq:eq2}
\end{eqnarray}
where the space scaling $dX=dx\sqrt{s}/a(x)$ was chosen to make the normalized coupling coefficient unity, and $\epsilon(x)=2 d(X\Delta q(X)/dX$ is the normalized detuning from the Bragg frequency. 

The wave, roughly speaking, reflects “from the bandgap”, i.e. from the position $X_0$ corresponding to the Bragg frequency, with $\epsilon(X_0)=0$. In general (for arbitrary chirp) Eq. (\ref{eq:eq2}) cannot be solved analytically. However, in a simple case when the normalized detuning varies linearly around zero $\epsilon(X)=\epsilon_1(X-X_0)$, Eq. (\ref{eq:eq2}) has an analytical solution in the form
\begin{eqnarray}
A(X)=c_1H_{\imath/\epsilon_1}(X\sqrt{\imath\epsilon_1/2}),
\label{eq:eq3}
\end{eqnarray}
where $H_n$ is the Hermite polynomial of imaginary order. The counter-propagating field obeys a similar expression. The integration constant $c_1=H_{\imath/\epsilon_1}(X_F\sqrt{\imath\epsilon_1/2})$ is determined by the boundary conditions, by imposing that the amplitude of the forward wave at the front face $X=X_F$ equals unity. $\epsilon_1=d\epsilon(X)/dX|_{X=X_F}$ or, in terms of initial variables, $\epsilon_1=4\pi\alpha/s$, which estimated for experimental parameters results $\epsilon_1=3$. 

\begin{figure}
\begin{center}
\includegraphics[width=85mm]{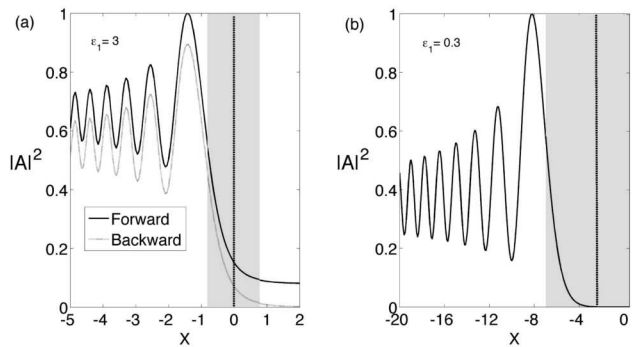}
\end{center} %\begin{quote}
\caption{(Color online) Solutions of (3), i.e. the intensity of the forward (continuous line) and backward (dashed line) field along the chirped structure. (a) (with  ) corresponds to experimental configuration, (b) (with  ) is shown for comparison, to illustrate soft reflections for a substantially smaller chirp. The vertical dashed lines indicate the center of the bandgap, and the shaded areas the bandgap itself.}
 \label{fig:fig5}
%\end{quote} 
\end{figure}

In Fig. \ref{fig:fig5} we present the amplitude of the acoustic intensity of the forward and backward waves for linearly chirped crystals as follows from Eq. (\ref{eq:eq3}). The acoustic field is nearly exponential in the bandgap, and oscillatory in front of it. The oscillations, with the period and amplitude increasing as the wave approaches the band-gap, are large-scale oscillations, which originate from the energy exchange between the forward and backward waves. These large-scale oscillations correspond to oscillations of the envelope of the Bloch modes observed in Fig. \ref{fig:fig4}, and are not due to conditions imposed at the entrance of the sonic crystal, e.g. some possible impedance mismatch. 

The controlled field enhancement effect is clearly visible again in Fig. \ref{fig:fig5}. From the analytical estimations in Eqs. (\ref{eq:eq1})-(\ref{eq:eq3}) and from Fig. \ref{fig:fig5} it follows that for maximal field enhancement of the wave intensity, the chirp must be as small as possible. For $\epsilon_1=0.3$ the maximum field enhancement could be around 6 times (in terms of intensities) if one compares the maximal and minimal values of the plot in Fig. \ref{fig:fig5}(b). In order to realize such enhancement the “entrance to the sonic crystal” must be placed to correspond to the deepest minimum of the solution (\ref{eq:eq3}), in this particular case at around the point $X=\simeq10$. For the parameters of the experiment, $\epsilon_1=0.3$, the enhancement of more than two times is predicted in this simplified approach. Also, as Fig. \ref{fig:fig5}(a) shows, a small portion of radiation is transmitted, i.e. “leaks” through the band gap. Such tunneling, analogous to Landau-Zenner tunneling, is due to slightly too fast (no more adiabatic) chirp.

Concluding, in this Letter we have predicted and experimentally demonstrated a novel mechanism for extraordinary sound field enhancement in a chirped crystals, specifically in chirped sonic crystals. The acoustic wave energy can be selectively concentrated at particular depth of the crystal depending on the frequency and on the parameters of the structure. At these bright planes, a substantial increase of the energy was recorded for linear chirp and for frequencies around the first gap along the X direction of structure. The experimental study was performed in a macroscopic sonic crystal irradiated by acoustic waves in audible regime, where the measurements in the interior of the crystal are possible.

We note, that the reported enhancement of the field is of different nature from the enhancement of the fields in resonant systems, e.g. in Fabry-P\'erot resonators, or in finite size periodic (unchirped) structures. The field inside the Fabry-P\'erot-like systems increases by resonator quality factor $Q$, however the enhancement occurs for a narrow frequency range (proportional to $Q^{-1}$). Therefore in average the intensity of broad frequency band radiation does not increase at all in Fabry-P\'erot-like resonant systems. The chirped sonic crystal studied here is not a resonant system, and the field enhancement occurs for broad frequency range (just the position of field localization is frequency dependent). 
   
A simple couple mode theory is proposed, that interprets and estimates the observed effects. The theory is based on generic equations describing wave propagation in periodically modulated media. The presented results have a universal character, i.e. valid irrespective to the nature of the waves. This implies that the reported effect, the controlled enhancement of the wave energy can be observed e.g. for optical waves, microwaves, or perhaps some exotic waves as surface plasmon waves.  

In the field of acoustics the results are independent of the spatial scale of the structure, and in principle the phenomenon could be scaled-down and observed in micro- or nano-scale phononic (so called hypersonic) crystals \cite{Gorishnyy05}. At these scales sound waves are described in terms of phonons, and the ideas presented in this work could find application for heat management in acoustical or acousto-optical devices. Recent works in this direction show indeed that manipulation of phonon dispersion properties can allow thermal transport control \cite{Hopkins11}. Generally, the effect of wave energy concentration demonstrated in present work, opens a possibility of increasing the efficiency of detectors and absorbers, both in acoustics and optics, since slow phonons and photons can be absorbed and harvested with a higher probability.

\begin{acknowledgments}
The work was supported by Spanish Ministry of Science and Innovation and European Union FEDER through projects FIS2011-29731-C02-01 and -02, also MAT2009-09438. V.R.G. is grateful for the support of post-doctoral contracts of the UPV CEI-01-11. K.S. acknowledges the grant of UPV PAID-02-01. We acknowledge the Centro de Tecnolog\'ias F\'isicas: Ac\'ustica, Materiales y Astrof\'isica and the Sonic Crystal Technologies Research Group of the Universitat Polit\`ecnica de Val\`encia for the use of the anechoic chamber and the 3DReAMS respectively.
\end{acknowledgments}

%\bibliography{Chirped}

%merlin.mbs apsrev4-1.bst 2010-07-25 4.21a (PWD, AO, DPC) hacked
%Control: key (0)
%Control: author (8) initials jnrlst
%Control: editor formatted (1) identically to author
%Control: production of article title (-1) disabled
%Control: page (0) single
%Control: year (1) truncated
%Control: production of eprint (0) enabled
%

\end{document}